\documentclass[conference,letterpaper]{IEEEtran}
\IEEEoverridecommandlockouts

\usepackage{cite}
\usepackage{amsmath,amssymb,amsfonts}
\usepackage{algorithmic}
\usepackage{graphicx}
\usepackage{textcomp}
\usepackage{xcolor}
\usepackage{balance}
\usepackage{url}
\usepackage{hyperref}
\usepackage{booktabs} 

\usepackage{xcolor}

\def\z{\phantom{0}}

\definecolor{numcolor}{RGB}{180, 0, 0}  
\newcommand{\num}[1]{{#1}}
\newcommand{\Fix}[1]{\textcolor{red}{[[#1]]}}
\newcommand{\CodeIn}[1]{{\texttt{#1}}}

\newcommand{\revisionTwo}[1]{{#1}}
\newcommand{\revision}[1]{{#1}}

\newcommand{\CodeUnderTest}{\textit{code under test}}
\newcommand{\Comment}[1]{}
\newcommand{\Space}[1]{}

\newcommand{\TotalJavaPRs}{\num{1278}}
\newcommand{\TotalPython}{\num{7191}}

\newcommand{\TotalJavaPRsAnalyzed}{\num{532}}
\newcommand{\TotalJavaProjectsAnalyzed}{\num{68}}

\newcommand{\TotalPythonPRsAnalyzed}{\num{4350}}
\newcommand{\TotalPythonProjectsAnalyzed}{\num{448}}

\newcommand{\JavaTotalBuild}{\num{16}}
\newcommand{\JavaTotalChore}{\num{17}}

\newcommand{\JavaTotalDocs}{\num{504}}
\newcommand{\JavaTotalFeat}{\num{180}}
\newcommand{\JavaTotalFix}{\num{277}}

\newcommand{\JavaTotalRefactor}{\num{53}}

\newcommand{\JavaTotalTest}{\num{210}}

\newcommand{\PythonTotalBuild}{\num{300}}
\newcommand{\PythonTotalChore}{\num{202}}
\newcommand{\PythonTotalCi}{\num{214}}
\newcommand{\PythonTotalDocs}{\num{1376}}
\newcommand{\PythonTotalFeat}{\num{2315}}
\newcommand{\PythonTotalFix}{\num{1802}}
\newcommand{\PythonTotalOther}{\num{6}}
\newcommand{\PythonTotalPerf}{\num{41}}
\newcommand{\PythonTotalRefactor}{\num{438}}
\newcommand{\PythonTotalRevert}{\num{2}}
\newcommand{\PythonTotalStyle}{\num{46}}
\newcommand{\PythonTotalTest}{\num{449}}

\newcommand{\TotalVennPRs}{\num{4882}} 
\newcommand{\TotalSourcePRs}{\num{4387}} 
\newcommand{\TotalSourcePRsPython}{\num{3857}}
\newcommand{\TotalSourcePRsJava}{\num{530}}
\newcommand{\VennSourceOnly}{\num{2211}} 
\newcommand{\SourceWithTests}{\num{2176}} 

\newcommand{\SourceWithoutTests}{\num{2211}} 

\newcommand{\TestsOnlyPRs}{\num{495}} 

\newcommand{\PRsAddingTests}{\num{1983}} 
\newcommand{\PRsModifyingTests}{\num{1500}} 
\newcommand{\PRsAddAndModify}{\num{812}} 

\newcommand{\PctSourceWithTests}{\num{49.6}} 
\newcommand{\PctSourceWithoutTests}{\num{50.4}} 

\newcommand{\PctTestsOnlyPRs}{\num{10.1}} 

\newcommand{\PctSourceWithoutTestsMutual}{\num{45.3}} 
\newcommand{\PctSourceWithTestsMutual}{\num{44.6}} 

\newcommand{\JavaFeatBefore}{39.7}
\newcommand{\JavaFeatAfter}{87.0}
\newcommand{\JavaFeatDelta}{+47.3}
\newcommand{\JavaFeatPctImp}{82.4}
\newcommand{\JavaFeatImp}{14}
\newcommand{\JavaFeatBuilt}{17}
\newcommand{\JavaFeatP}{< 0.001}
\newcommand{\JavaFeatSig}{***}
\newcommand{\JavaFixBefore}{74.8}
\newcommand{\JavaFixAfter}{79.7}
\newcommand{\JavaFixDelta}{+5.0}
\newcommand{\JavaFixPctImp}{27.6}
\newcommand{\JavaFixImp}{\z8}
\newcommand{\JavaFixBuilt}{29}
\newcommand{\JavaFixP}{0.006}
\newcommand{\JavaFixSig}{\z**}
\newcommand{\JavaTestBefore}{62.5}
\newcommand{\JavaTestAfter}{87.5}
\newcommand{\JavaTestDelta}{+25.0}
\newcommand{\JavaTestPctImp}{50.0}
\newcommand{\JavaTestImp}{\z1}
\newcommand{\JavaTestBuilt}{\z2}
\newcommand{\JavaTestP}{0.500}
\newcommand{\JavaTestSig}{n.s.}
\newcommand{\JavaDocsBefore}{97.1}
\newcommand{\JavaDocsAfter}{97.1}
\newcommand{\JavaDocsDelta}{+0.0}
\newcommand{\JavaDocsPctImp}{0.0}
\newcommand{\JavaDocsImp}{\z0}
\newcommand{\JavaDocsBuilt}{14}
\newcommand{\JavaDocsP}{n/a}
\newcommand{\JavaDocsSig}{n.s.}
\newcommand{\JavaPerfBefore}{92.9}
\newcommand{\JavaPerfAfter}{92.9}
\newcommand{\JavaPerfDelta}{+0.0}
\newcommand{\JavaPerfPctImp}{0.0}
\newcommand{\JavaPerfImp}{\z0}
\newcommand{\JavaPerfBuilt}{\z2}
\newcommand{\JavaPerfP}{n/a}
\newcommand{\JavaPerfSig}{n.s.}
\newcommand{\JavaOverallBefore}{70.5}
\newcommand{\JavaOverallAfter}{86.1}
\newcommand{\JavaOverallDelta}{+15.6}
\newcommand{\JavaOverallPctImp}{35.9}
\newcommand{\JavaOverallImp}{23}
\newcommand{\JavaOverallBuilt}{64}
\newcommand{\JavaOverallP}{< 0.001}
\newcommand{\JavaOverallSig}{***}
\newcommand{\PythonFeatBefore}{18.3}
\newcommand{\PythonFeatAfter}{31.3}
\newcommand{\PythonFeatDelta}{+13.0}
\newcommand{\PythonFeatPctImp}{27.2}
\newcommand{\PythonFeatImp}{99}
\newcommand{\PythonFeatBuilt}{364}
\newcommand{\PythonFeatP}{< 0.001}
\newcommand{\PythonFeatSig}{***}
\newcommand{\PythonFixBefore}{36.4}
\newcommand{\PythonFixAfter}{42.9}
\newcommand{\PythonFixDelta}{+6.6}
\newcommand{\PythonFixPctImp}{17.7}
\newcommand{\PythonFixImp}{25}
\newcommand{\PythonFixBuilt}{141}
\newcommand{\PythonFixP}{< 0.001}
\newcommand{\PythonFixSig}{***}
\newcommand{\PythonTestBefore}{25.7}
\newcommand{\PythonTestAfter}{23.5}
\newcommand{\PythonTestDelta}{-2.2}
\newcommand{\PythonTestPctImp}{8.3}
\newcommand{\PythonTestImp}{\z3}
\newcommand{\PythonTestBuilt}{\z36}
\newcommand{\PythonTestP}{0.633}
\newcommand{\PythonTestSig}{n.s.}
\newcommand{\PythonRefactorBefore}{42.9}
\newcommand{\PythonRefactorAfter}{46.6}
\newcommand{\PythonRefactorDelta}{+3.7}
\newcommand{\PythonRefactorPctImp}{11.4}
\newcommand{\PythonRefactorImp}{\z5}
\newcommand{\PythonRefactorBuilt}{\z44}
\newcommand{\PythonRefactorP}{0.088}
\newcommand{\PythonRefactorSig}{n.s.}
\newcommand{\PythonDocsBefore}{27.3}
\newcommand{\PythonDocsAfter}{33.5}
\newcommand{\PythonDocsDelta}{+6.2}
\newcommand{\PythonDocsPctImp}{21.4}
\newcommand{\PythonDocsImp}{\z3}
\newcommand{\PythonDocsBuilt}{\z14}
\newcommand{\PythonDocsP}{0.231}
\newcommand{\PythonDocsSig}{n.s.}
\newcommand{\PythonChoreBefore}{19.4}
\newcommand{\PythonChoreAfter}{19.4}
\newcommand{\PythonChoreDelta}{+0.0}
\newcommand{\PythonChorePctImp}{0.0}
\newcommand{\PythonChoreImp}{\z0}
\newcommand{\PythonChoreBuilt}{\z\z3}
\newcommand{\PythonChoreP}{n/a}
\newcommand{\PythonChoreSig}{n.s.}
\newcommand{\PythonPerfBefore}{0.0}
\newcommand{\PythonPerfAfter}{0.0}
\newcommand{\PythonPerfDelta}{+0.0}
\newcommand{\PythonPerfPctImp}{0.0}
\newcommand{\PythonPerfImp}{\z0}
\newcommand{\PythonPerfBuilt}{\z\z1}
\newcommand{\PythonPerfP}{n/a}
\newcommand{\PythonPerfSig}{n.s.}
\newcommand{\PythonStyleBefore}{1.0}
\newcommand{\PythonStyleAfter}{1.5}
\newcommand{\PythonStyleDelta}{+0.5}
\newcommand{\PythonStylePctImp}{100.0}
\newcommand{\PythonStyleImp}{\z1}
\newcommand{\PythonStyleBuilt}{\z\z1}
\newcommand{\PythonStyleP}{0.500}
\newcommand{\PythonStyleSig}{n.s.}
\newcommand{\PythonCiBefore}{0.0}
\newcommand{\PythonCiAfter}{0.0}
\newcommand{\PythonCiDelta}{+0.0}
\newcommand{\PythonCiPctImp}{0.0}
\newcommand{\PythonCiImp}{\z0}
\newcommand{\PythonCiBuilt}{\z\z1}
\newcommand{\PythonCiP}{n/a}
\newcommand{\PythonCiSig}{n.s.}
\newcommand{\PythonOverallBefore}{24.8}
\newcommand{\PythonOverallAfter}{34.5}
\newcommand{\PythonOverallDelta}{+9.6}
\newcommand{\PythonOverallPctImp}{22.5}
\newcommand{\PythonOverallImp}{136}
\newcommand{\PythonOverallBuilt}{605}
\newcommand{\PythonOverallP}{< 0.001}
\newcommand{\PythonOverallSig}{***}

\newcommand{\PythonMissCatMethodCallTotal}{4,516}

\newcommand{\PythonMissCatMethodCallMiss}{73.6}
\newcommand{\PythonMissCatAssignmentTotal}{9,472}

\newcommand{\PythonMissCatAssignmentMiss}{63.1}
\newcommand{\PythonMissCatIfTotal}{3,094}

\newcommand{\PythonMissCatIfMiss}{62.5}
\newcommand{\PythonMissCatForTotal}{710}

\newcommand{\PythonMissCatForMiss}{54.1}
\newcommand{\PythonMissCatWhileTotal}{94}

\newcommand{\PythonMissCatWhileMiss}{69.1}
\newcommand{\PythonMissCatReturnTotal}{2,189}

\newcommand{\PythonMissCatReturnMiss}{71.4}
\newcommand{\PythonMissCatTryCatchTotal}{1,356}

\newcommand{\PythonMissCatTryCatchMiss}{81.0}
\newcommand{\PythonMissCatThrowTotal}{536}

\newcommand{\PythonMissCatThrowMiss}{82.3}
\newcommand{\PythonMissCatBreakTotal}{27}

\newcommand{\PythonMissCatBreakMiss}{81.5}
\newcommand{\PythonMissCatContinueTotal}{204}

\newcommand{\PythonMissCatContinueMiss}{90.2}
\newcommand{\PythonMissCatSwitchTotal}{7}

\newcommand{\PythonMissCatSwitchMiss}{100.0}
\newcommand{\PythonMissCatDefinitionTotal}{3,062}

\newcommand{\PythonMissCatDefinitionMiss}{54.1}
\newcommand{\PythonMissCatImportTotal}{2,930}

\newcommand{\PythonMissCatImportMiss}{58.4}
\newcommand{\PythonMissCatOtherTotal}{1,171}

\newcommand{\PythonMissCatOtherMiss}{58.3}
\newcommand{\JavaMissCatMethodCallTotal}{403}

\newcommand{\JavaMissCatMethodCallMiss}{36.7}
\newcommand{\JavaMissCatAssignmentTotal}{671}

\newcommand{\JavaMissCatAssignmentMiss}{16.1}
\newcommand{\JavaMissCatIfTotal}{308}

\newcommand{\JavaMissCatIfMiss}{13.3}
\newcommand{\JavaMissCatForTotal}{31}

\newcommand{\JavaMissCatForMiss}{19.4}
\newcommand{\JavaMissCatWhileTotal}{21}

\newcommand{\JavaMissCatWhileMiss}{9.5}
\newcommand{\JavaMissCatReturnTotal}{367}

\newcommand{\JavaMissCatReturnMiss}{39.8}
\newcommand{\JavaMissCatTryCatchTotal}{43}

\newcommand{\JavaMissCatTryCatchMiss}{86.0}
\newcommand{\JavaMissCatThrowTotal}{80}

\newcommand{\JavaMissCatThrowMiss}{67.5}
\newcommand{\JavaMissCatBreakTotal}{3}

\newcommand{\JavaMissCatBreakMiss}{33.3}
\newcommand{\JavaMissCatContinueTotal}{2}

\newcommand{\JavaMissCatContinueMiss}{0.0}
\newcommand{\JavaMissCatSwitchTotal}{1}

\newcommand{\JavaMissCatSwitchMiss}{0.0}
\newcommand{\JavaMissCatDefinitionTotal}{35}

\newcommand{\JavaMissCatDefinitionMiss}{22.9}
\newcommand{\JavaMissCatOtherTotal}{103}

\newcommand{\JavaMissCatOtherMiss}{22.3}

\newcommand{\TotalPRAnalyzed}{4882}

\newcommand{\JavaExistingCov}{61.5}
\newcommand{\PythonExistingCov}{27.0}
\newcommand{\JavaPRBuilt}{213}
\newcommand{\PythonPRBuilt}{1664}

\begin{document}

\title{Test Coverage Analysis of Agentic Pull Requests}

\author{
\IEEEauthorblockN{Atish Kumar Dipongkor}
\IEEEauthorblockA{University of Central Florida \\
Orlando, FL, USA \\
atish.kumardipongkor@ucf.edu}
\and
\IEEEauthorblockN{Talank Baral}
\IEEEauthorblockA{George Mason University \\
Fairfax, VA, USA \\
tbaral@gmu.edu}
\and
\IEEEauthorblockN{Wing Lam}
\IEEEauthorblockA{George Mason University \\
Fairfax, VA, USA \\
winglam@gmu.edu}
\and
\IEEEauthorblockN{Kevin Moran}
\IEEEauthorblockA{University of Central Florida \\
Orlando, FL, USA \\
kpmoran@ucf.edu}
}

\maketitle

\begin{abstract}
AI coding agents increasingly submit complete pull requests (PRs) with minimal human intervention, shifting software development from AI-assisted to \textit{autonomous} workflows. As these agents become more prevalent, \revision{ensuring the code they generate is adequately tested, by existing tests or by tests the agents write, is critical to preventing regressions, yet little is known about testing in agentic PRs}. To address this gap, we analyze \num{\TotalPRAnalyzed} agent-generated PRs from the AIDev dataset (\num{\TotalJavaPRsAnalyzed} Java and \num{\TotalPythonPRsAnalyzed} Python PRs) produced by five coding agents. We study (i) how often agents include test changes and (ii) how well covered are code changes by existing and agent-written tests.
\revision{Agents include test changes in only \PctSourceWithTests{}\% of PRs that change \CodeUnderTest{} files. Existing tests provide an incomplete safety net: they cover \num{61.5}\% of agents' changed executable lines in Java and only \num{27.0}\% in Python, where \num{64.8}\% of PRs have no changed line executed by any existing test.\Space{\Fix{reword last part of last sentence}}
Agent-written tests improve coverage over existing tests, but only in a minority of PRs: \num{\JavaOverallPctImp}\% of Java and \num{\PythonOverallPctImp}\% of Python \emph{Code\,+\,Tests} PRs show a coverage gain.}
Across both languages, error-handling constructs (e.g., \CodeIn{try} and \CodeIn{catch} blocks) are the most consistently under-tested, with miss rates reaching \num{\JavaMissCatTryCatchMiss}\% in Java and \num{\PythonMissCatTryCatchMiss}\% in Python. 
These findings motivate coverage-aware development practices, coverage feedback loops for coding agents, and evaluation benchmarks that measure test quality to better help agents reliably test their own code.
\end{abstract}


\vspace{-0.5em}
\section{Introduction}
\label{sec:intro}

\renewcommand{\thefootnote}{\fnsymbol{footnote}}
\footnotetext[1]{This work is supported by the NSF CCF-2423813 \& CCF-2338287\Comment{Wing's CAREER} grants.}

The emergence of autonomous AI coding agents marks a paradigm shift in software engineering practices~\cite{li2025aidev}. As agents autonomously carry out software development tasks at scale, ensuring the quality of generated code is becoming a notable bottleneck, which raises fundamental questions about software quality assurance practices, particularly regarding testing.

Testing serves as one of the most important tools for verifying functional correctness and preventing
regressions~\cite{fraser2011evosuite,ammann2016introduction,pezze2007software,swebok2014,graves2001empirical}.
However, how well tested the code introduced by autonomous coding agents is remains unknown. Agent-introduced changes can be exercised by two sources of tests: the project's existing tests, or tests the agent writes alongside its changes. Neither source has been evaluated in practice: benchmarks like SWE-bench~\cite{jimenez2024swebench} assess agents' ability to resolve issues by passing existing tests, but do not measure whether those tests exercise the code agents introduce, nor whether agents create adequate tests for their new functionality. Prior research on AI-based test generation does assess test adequacy, demonstrating that Large Language Models (LLMs)~\cite{chen2021evaluating} can produce syntactically valid tests~\cite{mastropaolo2023robustness, schafer2024empirical} and achieve reasonable code coverage when guided by appropriate prompting strategies~\cite{lemieux2023coverup, ryan2024code}. E.g., Meta's
TestGen-LLM~\cite{meta2024testgen} and benchmarks, such as
SWT-Bench~\cite{mundler2024swtbench}, have shown promise in augmenting existing tests with LLM-generated tests that improve coverage or reproduce bugs. However, these studies evaluate settings where AI is explicitly instructed to generate tests. Such controlled evaluations do not reveal how well agents test their own changes when operating autonomously.

Recent empirical studies of agentic coding reveal concerning patterns: agents often prioritize functional code over comprehensive testing, and nearly one-third of merged agent-generated pull requests (PRs) require subsequent bug fixes or refactoring~\cite{watanabe2025use}. 
Similarly, surveys indicate that while developers use AI assistants for test generation, developers also express skepticism about the quality and coverage of AI-generated tests~\cite{liang2024ai_assistants}. These findings suggest that while agents can generate tests on demand, they may neglect to test the new code paths introduced during autonomous development. 

This disconnect has substantial implications for software maintenance. When agents introduce untested code paths, they create technical debt~\cite{cunningham1992wycash,sculley2015hidden, bhatia2024empirical} that accumulates silently until regression failures emerge in production~\cite{pecorelli2021relation}. Understanding how agents currently approach testing and where they fall short is essential for establishing quality gates in AI-augmented development pipelines and for guiding the evolution of more testing-conscious agents.

In this paper, we present emerging results from \revision{an} empirical analysis into how well-tested code generated by AI agents is in practice\Space{\Fix{ Didn't SWT-bench or others already analyze how well tested code generated by AI agents is? What are we ``first'' at doing and is it expressed in this sentence? If not, replace ``the first'' to ``an''}}. 
We analyze \TotalJavaPRsAnalyzed{} Java and \TotalPythonPRsAnalyzed{} Python pull requests from the AIDev dataset~\cite{li2025aidev}, spanning five prominent AI coding agents\Space{: Claude Code~\cite{anthropic2025claudecode}, Codex~\cite{openai2021codex}, Copilot~\cite{github2022copilot}, Cursor~\cite{cursor2024}, and Devin~\cite{cognition2024devin}}. 
Our analysis addresses two main research questions:


\noindent\textbf{RQ$_1$: How often do agentic PRs contain test code changes\Space{\Fix{ code change}s}?} 
This RQ establishes a baseline of testing behavior. When agents create
pull requests, do they include \textit{test changes} (adding new/modifying existing tests) alongside code changes? We classify PRs into three categories: those that change \textit{\textbf{both}} \CodeUnderTest{} and \textit{test} files, those that change \textit{\textbf{only}} \CodeUnderTest{}, and those that change \textit{\textbf{only}} \textit{test} files. Findings for this RQ
show the ``broad strokes'' of agentic behavior.

\par
\noindent\textbf{RQ$_2$: What is the test coverage of code changes in agentic PRs?} Beyond presence, we assess test effectiveness through coverage analysis, across two sources of tests:

\noindent\textbf{RQ$_{2a}$ (Coverage by existing tests)}: \textit{What is the coverage of
agent-written code for existing tests?} We analyze all merged PRs
that change code under test, i.e., both Code-Under-Test-Only and
Code\,+\,Tests PRs. For the latter source, we measure coverage with the
agent's test changes removed.\Space{ Agents may rely on existing tests or
add untested code.}

\noindent\textbf{RQ$_{2b}$ (Coverage gain by agent-written tests)}:
\textit{What is the coverage gain of agent-written code when agent-written tests are combined with existing tests?} We analyze Code\,+\,Tests PRs, comparing
coverage with and without the agent's test changes. High gains
indicate targeted, effective tests; low gains expose superficial
testing.


    

    

Our analysis reveals four main findings:
(i) \PctSourceWithoutTests\% of agentic PRs that modify \CodeUnderTest{} include no test changes at all. (ii) \revision{Existing tests do not fully cover agent-written code: only \JavaExistingCov{}\% of agents' changed executable lines in Java and \PythonExistingCov{}\% in Python are covered,  and 64.8\% of Python PRs have none of their changed lines executed by any existing test.\Space{\Fix{reword last part of last sentence}}} (iii) \revision{Agent-written tests significantly improve coverage of the agents' changes on average (+15.6 pp in Java, +9.6 pp in Python) over existing tests, but the gains are concentrated in a minority of PRs: only \num{\JavaOverallPctImp}\% of Java and \num{\PythonOverallPctImp}\% of Python Code + Tests PRs show any gain\Space{ beyond the existing tests}\Space{\Fix{globally refer to these tests as just ``existing tests''. It's confusing to call the same thing in multiple ways.}}}. (iv) Error-handling constructs are\Space{ systematically} under-tested in both languages, with miss rates of \num{\JavaMissCatTryCatchMiss}\% in Java and \num{\PythonMissCatTryCatchMiss}\% in Python. Our results provide actionable insights for practitioners integrating AI agents into their workflows and for improving agent capabilities (Section~\ref{sec:discussion}). Our results also illustrate notable shortcomings related to the ``tested-ness'' of agent generated code, signaling the need for new research\Space{ directions} that enable agents to better check and test their work.

This paper makes the following contributions:
\begin{itemize}
    \item The first cross-language empirical study of testing behavior in real-world agentic PRs, spanning \num{\TotalPRAnalyzed} PRs across five AI coding agents and two languages.
    \item A coverage analysis of agent-generated code for existing tests and existing tests combined with agent-written tests.
    \item Identification of specific weak spots in agentic testing—error-handling constructs in both languages and Python's near-zero existing-test fallback—that point to concrete priorities for future agent design and continuous integration (CI)\Space{\Fix{undefined}} gating.
    \item A publicly available dataset~\cite{anonymous2026aiprtest,dipongkor2026artifact} of  PR patches, test-only patches, coverage reports,
and added lines labeled by syntactic construct (e.g.,
\CodeIn{Assignment}, \CodeIn{Return}, \CodeIn{Try-Catch}) to support
future research.
\end{itemize}

\section{Dataset}
\label{sec:dataset}




Java (\TotalJavaPRs{}) and Python (\TotalPython{}) PRs are taken from the AIDev dataset version 3~\cite{li2025aidev}, spanning five agents across twelve conventional commit categories. Codex is the most popular agent in both languages, contributing 1018 Java and 5209 Python PRs; Copilot, Cursor, Claude Code, and Devin contribute the remainder. The category mix differs by language: Java is documentation-heavy (\texttt{\textbf{\small docs}} \JavaTotalDocs{}, \texttt{\textbf{\small fix}} \JavaTotalFix{}, \texttt{\textbf{\small test}} \JavaTotalTest{}, \texttt{\textbf{\small feat}} \JavaTotalFeat{}, \texttt{\textbf{\small refactor}} \JavaTotalRefactor{}, \texttt{\textbf{\small chore}} \JavaTotalChore{}, \texttt{\textbf{\small build}} \JavaTotalBuild{}), while Python is feature-heavy (\texttt{\textbf{\small feat}} \PythonTotalFeat{}, \texttt{\textbf{\small fix}} \PythonTotalFix{}, \texttt{\textbf{\small docs}} \PythonTotalDocs{}, \texttt{\textbf{\small test}} \PythonTotalTest{}, \texttt{\textbf{\small refactor}} \PythonTotalRefactor{}, \texttt{\textbf{\small build}} \PythonTotalBuild{}, \texttt{\textbf{\small ci}} \PythonTotalCi{}, \texttt{\textbf{\small chore}} \PythonTotalChore{}, \texttt{\textbf{\small style}} \PythonTotalStyle{}, \texttt{\textbf{\small perf}} \PythonTotalPerf{}, \texttt{\textbf{\small other}} \PythonTotalOther{}, \texttt{\textbf{\small revert}} \PythonTotalRevert{}). To answer our RQs, we apply two filters to isolate PRs that modify executable code. First, we retain PRs that change at least one .java or .py file. Second, we parse the changed files with tree-sitter~\cite{treesitter2018} and discard PRs whose modifications fall entirely within comments or docstrings. 
These filters yield \TotalJavaPRsAnalyzed{} Java PRs across \TotalJavaProjectsAnalyzed{} repositories and \TotalPythonPRsAnalyzed{} Python PRs across \TotalPythonProjectsAnalyzed{} repositories, which we use for RQ$_1$. For the coverage analysis in RQ$_2$, we consider only merged PRs from repositories with at least 10 agentic PRs, discarding\Comment{ toy \Fix{don't say they are ``toy''. it's unclear what that means and we don't really know if they are ``toy'' or not anyway}} projects and repositories with too few agentic PRs to analyze reliably. 
This filter retains \num{14} Java and \num{55} Python repositories. 
The number of PRs used in RQ$_2$ is further discussed in Section~\ref{sec:artifact-gen}.\Comment{
This filter retains \Fix{? Java PRs across} 14 repositories and \Fix{? Python PRs across} 55 repositories. \Fix{if you are short on space, then remove the count of repositories. you should have count of PRs as that is actually what matters.}}

\section{Methodology}
\label{sec:methodology}

\subsection{Artifact Generation}\label{sec:artifact-gen}
\textbf{PR patch.} The AIDev dataset does not provide PR-level patches. We cloned each repository locally and reconstructed the unified diff for every PR using its base and head SHAs (\texttt{\textbf{\small git diff <base>..<head>}}), yielding a self-contained patch per PR.\Comment{\Fix{already described in Section 2}We collect \TotalJavaPRsAnalyzed{} Java and \TotalPythonPRsAnalyzed{} Python patches, and use them in $RQ_{1}$.}

\textbf{Extracting test-only code from the patches.} From each PR patch, we extract the changes corresponding to test files\Space{ if any}. The test-only patch enables us to remove those tests selectively during coverage measurement.

\textbf{PR-level coverage.} For each PR, we execute the repository's entire test suite, collecting line coverage with JaCoCo~\cite{jacoco} for Java and \texttt{pytest-cov}~\cite{pytest-cov} for Python. Of the repositories in the coverage subset (Section II), 10 of the 14 Java and 34 of the 55 Python repositories could be built and instrumented, yielding coverage results for 213 of the 532 Java PRs and 1664 of the 4350 Python PRs.

\textbf{PR-level coverage (without tests).} For each PR with a non-empty test-only patch, we reverse-apply the test-only patch (\CodeIn{git\allowbreak{} apply\allowbreak{} -R}) on the PR's \texttt{\textbf{\small<head>}} commit and re-run the suite. The difference between the two runs isolates the coverage contributed by agent-written tests. This paired analysis is applicable to the 64 Java and 605 Python PRs (of the 213 and 1664 merged PRs with coverage results) that contain test changes; the remaining 149 Java and 1059 Python PRs modify only code under test and are analyzed with developer's existing tests.

\textbf{Identifying added \CodeUnderTest{} lines.} For each PR patch, we\Comment{ walk the patch line by line and} record, for every \CodeUnderTest{} file, the line numbers of all added (\texttt{+}) lines as they appear in the head-commit version of the file. Test files are excluded by directory (\texttt{\textbf{\small test/tests)}} and by language-specific filename conventions: JUnit/Spring/Failsafe patterns (\texttt{\textbf{\small *Test}}, \texttt{\textbf{\small *Tests}}, \texttt{\textbf{\small *TestCase}}, \texttt{\textbf{\small Test*}}, \texttt{\textbf{\small *IT}}, \texttt{\textbf{\small *ITCase}}) for Java, and PyTest/unit test patterns (\texttt{\textbf{\small test\_*.py}}, \texttt{\textbf{\small *\_test.py}}) for Python.

\textbf{Classifying added lines.} For each PR, we snapshot every \CodeUnderTest{} file at the PR's head commit and parse it with \texttt{\textbf{\small srcml --position}}~\revision{\cite{collard2013srcml}}\Space{ \Fix{cite}}. Following Zhu et al.~\cite{zhu2015analysis}, each added line is assigned the local tag of the deepest srcML element whose position range encloses its trimmed content (e.g., \texttt{\textbf{\small Assignment}}, \texttt{\textbf{\small If}}, \texttt{\textbf{\small Return}}). This typed inventory of added lines lets us characterize \emph{which syntactic constructs} agents systematically leave untested.

\subsection{RQ$_1$: Classifying Test Behavior}\label{sec:methodology:rq1}

We categorize each PR by the type of test changes, using two criteria derived from the PR patch: (i) whether a hunk changes a test file (identified by the rules in Section~\ref{sec:artifact-gen}), and (ii) whether a hunk adds a new test.

\textbf{Detecting newly added tests.} For each hunk, we compute the set of test-method declarations in the post-change version of the hunk (i.e., the file content at the PR's head commit) minus those in the pre-change version (i.e., the content at the base commit); the difference identifies tests \emph{introduced}, not merely edited, by the PR. A method counts as an added test if, in Java, it carries one of \texttt{\textbf{\small @Test}}, \texttt{\textbf{\small @ParameterizedTest}}, \texttt{\textbf{\small @RepeatedTest}}, \texttt{\textbf{\small @TestFactory}}, or \texttt{\textbf{\small @TestTemplate}} (JUnit~4/5, TestNG), or its name follows the JUnit~3 test name\Comment{\texttt{\textbf{\small testFoo}}} convention; in Python, a test is a \texttt{\textbf{\small def}} or \texttt{\textbf{\small async def}} whose method name begins with \texttt{\textbf{\small test}}, covering both \texttt{\textbf{\small pytest}} (\texttt{\textbf{\small test\_*}}) and \texttt{\textbf{\small unittest}} (\texttt{\textbf{\small test*}}). 
Body-only edits and annotation-only changes leave the declared-name set unchanged and are not counted as added test.

\textbf{PR classification.} Using criterion (i), we assign each PR to exactly one of three categories: Code + Tests PRs, which change both code under test and test files; Tests-only PRs, which change test files only; and Code-Under-Test-Only PRs, which change code under test with no test-file changes. The distribution of these categories is in Section~\ref{sec:rq1-res}\Space{IV-A \Fix{use ref instead}}. Using criterion (ii), a PR \textit{Adds Tests} if it introduces at least one new test, and \textit{Modifies Tests} if it changes test files without introducing a new test. A PR may both add and modify tests. Figure~\ref{fig:pr-classification} reports the category overlap.

\subsection{RQ2: Measuring Diff Coverage and Analysis}

To assess test effectiveness, we measure \textit{diff coverage}: the percentage of added code lines that are executed by tests.

\textbf{Coverage computation.} For each modified \CodeUnderTest{} file, we check every added line against the JaCoCo or \texttt{\textbf{\small pytest-cov}} report and label it as covered (at least one test executed it) or missed. File-level diff coverage is then computed as:
\vspace{-2ex}
\begin{equation}
\text{Diff Coverage}_{\text{file}} = \frac{|\text{Covered Lines}|}{|\text{Covered Lines}| + |\text{Missed Lines}|}
\end{equation}

\textbf{Aggregation.} We aggregate file-level diff coverage to the PR level by summing covered and missed lines across all changed \CodeUnderTest{} files in the PR, then applying the same ratio. Doing so nets a single diff-coverage value per PR. 

\emph{For Code + Tests PRs (RQ$_{2a}$}), we additionally compute PR-level diff coverage on the without-tests run (Section~\ref{sec:artifact-gen}), denoting the two values as $\text{DiffCov}_{\text{with}}$ and $\text{DiffCov}_{\text{without}}$. We report their difference,
\vspace{-1ex}
\begin{equation}
\Delta\text{DiffCov} = \text{DiffCov}_{\text{with}} - \text{DiffCov}_{\text{without}},
\end{equation}
\vspace{-1ex}
as the diff-coverage gain attributable to the agent's own tests.

\section{Results}
\label{sec:results}

\begin{figure}[tb]
    \centering
    \includegraphics[width=.7\columnwidth]{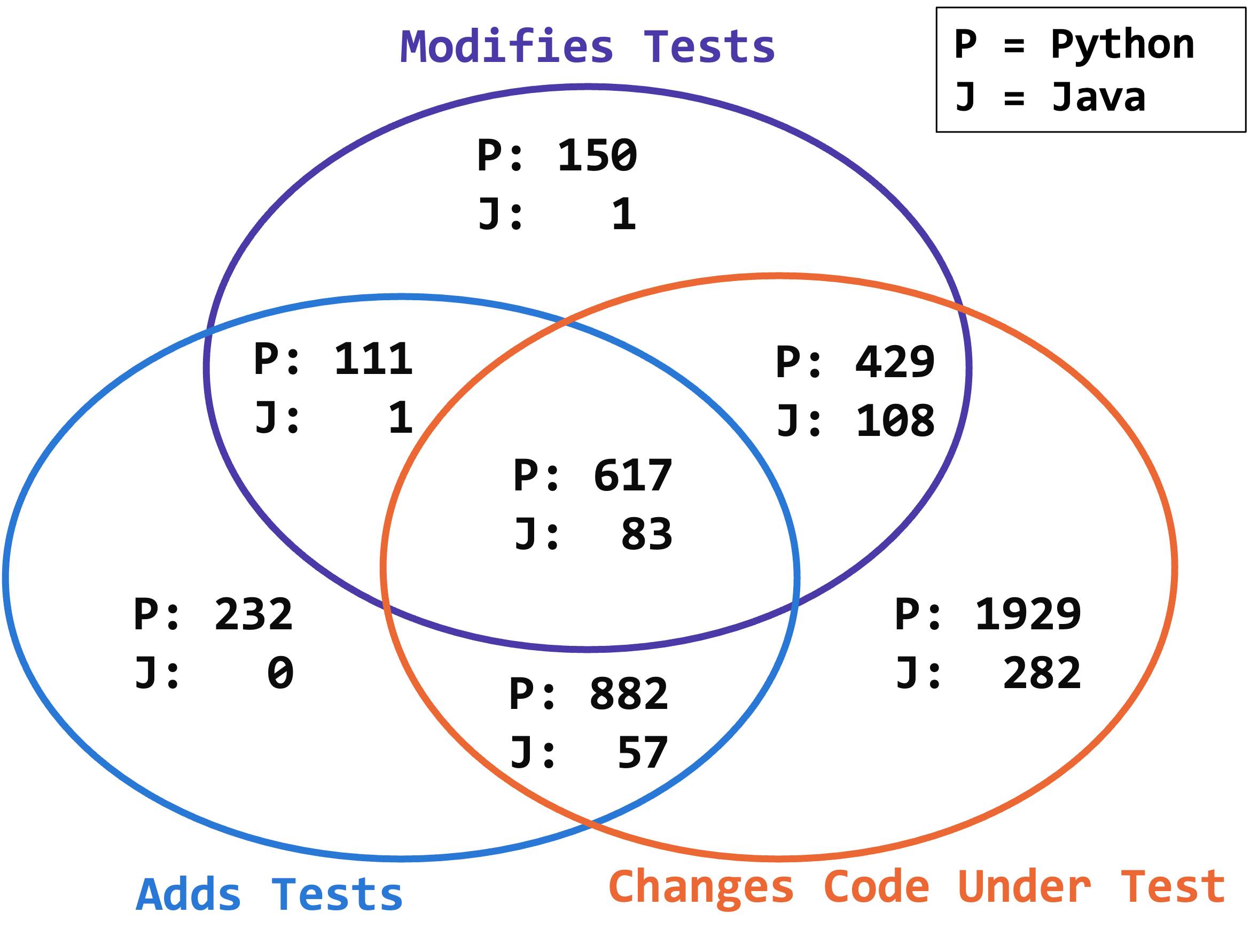}
    \vspace{-2.5ex}
    \caption{PR classification results. Circles denote non-exclusive PR attributes; each region shows the number of PRs.} 
    \vspace{-.5em}
    \vspace{.5ex}
    \label{fig:pr-classification}
\end{figure}

\subsection{RQ1 Results: Test Inclusion in Agentic PRs}
\label{sec:rq1-res}

Figure~\ref{fig:pr-classification} presents a Venn diagram of the intersection of PRs with three attributes: those that (i) add new tests, (ii) modify existing tests, and (iii) change only \CodeUnderTest{} files.

\textbf{Overall test inclusion.} 
Of the \TotalPRAnalyzed{} PRs in our dataset, \TotalSourcePRs{} modify \CodeUnderTest{} files (\TotalSourcePRsJava{} Java and \TotalSourcePRsPython{} Python); the remaining \TestsOnlyPRs{}{} change test files only, of which \num{430} are merged. 
Among the \TotalSourcePRs{} PRs that modify \CodeUnderTest{} files, \SourceWithTests{} (\PctSourceWithTests{}\%)
include test changes, of which \num{1346} are merged, while \SourceWithoutTests{} (\PctSourceWithoutTests{}\%) have no such changes, of which \num{1692} are merged.

\Space{
\textbf{PR Category Distribution.} We identify three categories:\Fix{Can we remove this section? The only info it adds is \# of merged PRs but is that relevant to the paper?} \Fix{Atish: one of the reviewers asked for it}
\begin{itemize}
    \item \textbf{Code + Tests PRs:} \SourceWithTests{} out of \TotalVennPRs{} PRs (\PctSourceWithTestsMutual{}\%) change both \CodeUnderTest{} and tests, of which 1346 are merged.
    \item \textbf{Tests-only PRs:} \TestsOnlyPRs{} out of \TotalVennPRs{} PRs (\PctTestsOnlyPRs{}\%) change tests without changing \CodeUnderTest{} lines, of which 430 are merged. 
    \item \textbf{Code-Under-Test-Only PRs:} \VennSourceOnly{} out of \TotalVennPRs{} PRs (\PctSourceWithoutTestsMutual{}\%) change \CodeUnderTest{} files only, of which 1692 are merged.\Space{without any test changes:}
\end{itemize}
}

\textbf{Test addition vs. modification.} When agents do include tests, they more frequently add new tests than modify existing ones. A total of \PRsAddingTests{} PRs add new tests, while \PRsModifyingTests{} PRs modify existing tests. \PRsAddAndModify{} PRs both add and modify tests.

\textbf{PR categories without test changes.}

As a majority of PRs (\num{53.0}\% Java, \num{44.3}\% Python)\Space{\Fix{globally search Java/Python and make sure Java comes before Python. Most of the paper seem to talk about Java before Python but every now and then like here, we have Python first.}} modify \CodeUnderTest{} files without any test changes, we examine their categories. Java is dominated by \texttt{\textbf{\small docs}} (\num{37.6}\%),  \texttt{\textbf{\small fix}} (\num{34.4}\%), and \texttt{\textbf{\small feat}} (\num{15.2}\%); Python by \texttt{\textbf{\small feat}} (\num{40.5}\%), \texttt{\textbf{\small fix}} (\num{30.3}\%), and \texttt{\textbf{\small refactor}} (\num{11.8}\%).

\begin{figure}[tb]
    \centering
    \includegraphics[width=0.8\linewidth]{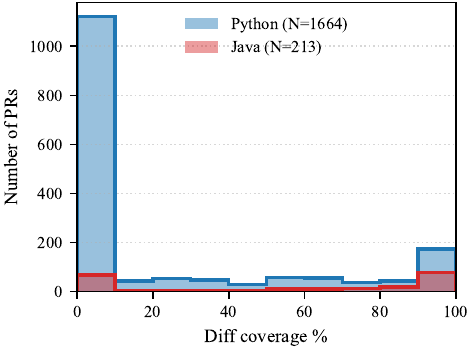}
    \vspace{-2ex}
    \caption{\revision{Existing-test diff coverage per merged PR.}}
    \label{fig:diff-cov-code-only}
    \vspace{-1em}
\end{figure}

\subsection{RQ2 Results: Test Coverage in Agentic PRs}
In this section, the results are from merged PRs (described in sections~\ref{sec:dataset} and~\ref{sec:artifact-gen}) for both languages.
\Space{\Fix{Make sure to review this section and conclusion to properly discuss the problem (agent generated code being covered by existing tests AND existing tests+agent tests), and the actual results should be presented in this order}}

\begin{table*}[tb]
\centering
\vspace{-1em}
\setlength{\tabcolsep}{2pt}
\caption{$\text{DiffCov}_{\text{without}}$ and $\text{DiffCov}_{\text{with}}$ by PR category for Code\,+\,Tests PRs, compared between Java and Python. Significance is assessed with the one-sided Wilcoxon signed-rank test~\cite{wilcoxon1992individual} ($H_1$: $\text{DiffCov}_{\text{with}}$ $>$ $\text{DiffCov}_{\text{without}}$).}
\label{tab:diff-coverage-by-category}
\vspace{-1em}
\begin{tabular}{l rrrrr rrrrr}
\toprule
& \multicolumn{5}{c}{\textbf{Java}} & \multicolumn{5}{c}{\textbf{Python}} \\
\cmidrule(lr){2-6} \cmidrule(lr){7-11}
\textbf{Category} & \textbf{$\text{DiffCov}_{\text{without}}$} & \textbf{$\text{DiffCov}_{\text{with}}$} & $\boldsymbol{\Delta}$ \textbf{(DiffCov)} & \textbf{\% Improved} & \textbf{$p$-value} & \textbf{$\text{DiffCov}_{\text{without}}$} & \textbf{$\text{DiffCov}_{\text{with}}$} & $\boldsymbol{\Delta}$ \textbf{(DiffCov)} & \textbf{\% Improved} & \textbf{$p$-value} \\
\midrule
feat & \JavaFeatBefore\% & \JavaFeatAfter\% & \JavaFeatDelta & \JavaFeatPctImp\% (\JavaFeatImp/\JavaFeatBuilt) & $\JavaFeatP$\textsuperscript{\JavaFeatSig} & \PythonFeatBefore\% & \PythonFeatAfter\% & \PythonFeatDelta & \PythonFeatPctImp\% (\PythonFeatImp/\PythonFeatBuilt) & $\PythonFeatP$\textsuperscript{\PythonFeatSig} \\
fix & \JavaFixBefore\% & \JavaFixAfter\% & \JavaFixDelta & \JavaFixPctImp\% (\JavaFixImp/\JavaFixBuilt) & $\JavaFixP$\textsuperscript{\JavaFixSig} & \PythonFixBefore\% & \PythonFixAfter\% & \PythonFixDelta & \PythonFixPctImp\% (\PythonFixImp/\PythonFixBuilt) & $\PythonFixP$\textsuperscript{\PythonFixSig} \\
test & \JavaTestBefore\% & \JavaTestAfter\% & \JavaTestDelta & \JavaTestPctImp\% (\JavaTestImp/\JavaTestBuilt) & $\JavaTestP$\textsuperscript{\JavaTestSig} & \PythonTestBefore\% & \PythonTestAfter\% & \PythonTestDelta & \PythonTestPctImp\% (\PythonTestImp/\PythonTestBuilt) & $\PythonTestP$\textsuperscript{\PythonTestSig} \\
refactor & — & — & — & — (\z0/\z0) & — & \PythonRefactorBefore\% & \PythonRefactorAfter\% & \PythonRefactorDelta & \PythonRefactorPctImp\% (\PythonRefactorImp/\PythonRefactorBuilt) & $\PythonRefactorP$\textsuperscript{\PythonRefactorSig} \\
docs & \JavaDocsBefore\% & \JavaDocsAfter\% & \JavaDocsDelta & \JavaDocsPctImp\% (\JavaDocsImp/\JavaDocsBuilt) & $\JavaDocsP$\textsuperscript{\JavaDocsSig} & \PythonDocsBefore\% & \PythonDocsAfter\% & \PythonDocsDelta & \PythonDocsPctImp\% (\PythonDocsImp/\PythonDocsBuilt) & $\PythonDocsP$\textsuperscript{\PythonDocsSig} \\
build & — & — & — & — (\z0/\z0) & — & — & — & — & — (\z0/\z\z0) & — \\
chore & — & — & — & — (\z0/\z0) & — & \PythonChoreBefore\% & \PythonChoreAfter\% & \PythonChoreDelta & \PythonChorePctImp\% (\PythonChoreImp/\PythonChoreBuilt) & $\PythonChoreP$\textsuperscript{\PythonChoreSig} \\
perf & \JavaPerfBefore\% & \JavaPerfAfter\% & \JavaPerfDelta & \JavaPerfPctImp\% (\JavaPerfImp/\JavaPerfBuilt) & $\JavaPerfP$\textsuperscript{\JavaPerfSig} & \PythonPerfBefore\% & \PythonPerfAfter\% & \PythonPerfDelta & \PythonPerfPctImp\% (\PythonPerfImp/\PythonPerfBuilt) & $\PythonPerfP$\textsuperscript{\PythonPerfSig} \\
style & — & — & — & — (\z0/\z0) & — & \PythonStyleBefore\% & \PythonStyleAfter\% & \PythonStyleDelta & \PythonStylePctImp\% (\PythonStyleImp/\PythonStyleBuilt) & $\PythonStyleP$\textsuperscript{\PythonStyleSig} \\
ci & — & — & — & — (\z0/\z0) & — & \PythonCiBefore\% & \PythonCiAfter\% & \PythonCiDelta & \PythonCiPctImp\% (\PythonCiImp/\PythonCiBuilt) & $\PythonCiP$\textsuperscript{\PythonCiSig} \\
\midrule
\textbf{Overall} & \textbf{\JavaOverallBefore\%} & \textbf{\JavaOverallAfter\%} & \textbf{\JavaOverallDelta} & \textbf{\JavaOverallPctImp\% (\JavaOverallImp/\JavaOverallBuilt)} & \textbf{$\JavaOverallP$\textsuperscript{\JavaOverallSig}} & \textbf{\PythonOverallBefore\%} & \textbf{\PythonOverallAfter\%} & \textbf{\PythonOverallDelta} & \textbf{\PythonOverallPctImp\% (\PythonOverallImp/\PythonOverallBuilt)} & \textbf{$\PythonOverallP$\textsuperscript{\PythonOverallSig}} \\
\bottomrule
\end{tabular}
\\[2pt]
\footnotesize Significance: \textsuperscript{***}$p<0.001$; \textsuperscript{**}$p<0.01$; \textsuperscript{*}$p<0.05$; n.s. = not significant.
\vspace{-1.3em}
\end{table*}

\noindent\textbf{RQ$_{2a}$: \revision{What is the coverage of agent-written code for existing tests}?} \revision{For all merged PRs that change \CodeUnderTest{}, we measure the fraction of added executable lines covered by the repository's existing tests. For Code\,+\,Tests PRs, we measure coverage with the agent's test changes removed (Section~\ref{sec:methodology}). This measurement indicates how much of the agent's diff the existing tests actually covers. Across the \JavaPRBuilt{} Java and \PythonPRBuilt{} Python merged PRs, existing tests cover \JavaExistingCov\% of changed executable lines in Java and only \PythonExistingCov\% in Python.}

\textbf{PR-level coverage.} 
\revision{The two languages exhibit sharply different coverage from existing tests. In Java, the median PR
has \num{71.1}\% of its changed executable lines covered by the existing tests, and \num{34.3}\% (\num{73}/\num{213}) are fully covered. Python tells a different story: the median PR has
\emph{zero} coverage, and existing tests cover none of the changed executable lines in \num{64.8}\% (\num{1079}/\num{1664}) of Python PRs. Figure~\ref{fig:diff-cov-code-only} shows the underlying distribution: both languages are bimodal with mass at 0\% and 100\%, but Python concentrates overwhelmingly at the zero-coverage end while Java is more evenly distributed.}

\textbf{File-level coverage.} \revision{\Space{The same asymmetry holds at the file level. }Across all changed \CodeUnderTest{} files with at least one executable line in the diff, aggregate diff coverage is \num{61.5}\% in Java (\num{259} files) versus \num{27.0}\% in Python (\num{2696} files). In Java, \num{50.2}\% of files are fully covered and only \num{18.5}\% are uncovered; in Python, the proportions invert: \num{24.5}\% are fully covered and \num{54.2}\% of files are uncovered.}

\noindent\textbf{Takeaway.} \revision{Existing tests are an incomplete safety net in both languages, and far weaker in Python: such tests cover \num{61.5}\% of changed lines in Java, but only \num{27.0}\% in Python. Existing tests also cover no changed lines in \num{64.8}\% of Python PRs.}

\noindent\textbf{RQ$_{2b}$: \revision{What is the coverage gain of agent-written code when agent-written tests are combined with existing tests}?} Table~\ref{tab:diff-coverage-by-category} reports $\text{DiffCov}_{\text{without}}$, $\text{DiffCov}_{\text{with}}$, and their delta across PR categories for Code\,+\,Tests PRs. 
Agent-written tests yield a statistically significant improvement: Java improves from \JavaOverallBefore{}\% to \JavaOverallAfter{}\% ($\Delta = \JavaOverallDelta$\,pp; $p < 0.001$) and Python from \PythonOverallBefore{}\% to \PythonOverallAfter{}\% ($\Delta = \PythonOverallDelta$\,pp; $p < 0.001$). The gain is concentrated in feature-implementing PRs, where agent tests deliver the largest deltas in both languages (Java: $\JavaFeatDelta$\,pp on \JavaFeatBuilt{} PRs; Python: $\PythonFeatDelta$\,pp on \PythonFeatBuilt{} PRs), followed by smaller but significant gains for bug fixes ($\JavaFixDelta$\,pp Java, $\PythonFixDelta$\,pp Python). Other categories—\texttt{\textbf{\small docs}}, \texttt{\textbf{\small refactor}}, \texttt{\textbf{\small chore}}, \texttt{\textbf{\small perf}}, \texttt{\textbf{\small style}}, \texttt{\textbf{\small ci}}—show little gain.

\noindent\textbf{Improvement is concentrated in a minority of PRs.} Only \JavaOverallPctImp{}\% of Java (\JavaOverallImp{}/\JavaOverallBuilt{}) and \PythonOverallPctImp{}\% of Python (\PythonOverallImp{}/\PythonOverallBuilt{}) Code\,+\,Tests PRs actually show any improvement. 
For the remainder, agent-supplied tests do not raise diff coverage. Inspecting these non-improving PRs reveals that the underlying causes differ sharply between the two languages. In Java, \num{\revisionTwo{42.2}}{\%} already reach \num{100}{\%} diff coverage from \revisionTwo{the existing tests}.\Space{\Fix{Does without-tests run mean test run with only existing tests? If so, say that.}}

Among the rest, agents delete more tests than they add (82 deleted vs.\ 31 added, a 2.6$\times$ ratio), \revisionTwo{with
another \num{51.2}\% editing only the bodies of existing tests -- such edits modify existing test behavior but introduce no new tests targeting the changed code}.\Space{ \Fix{What does body-only edits mean? Edits only within test method body? Such edits still change the tests but it doesn't change the number/names of tests. Do we mean number/names of tests when we say ``declared-test set unchanged''?}.} 
In Python, the pattern inverts: only \num{8}{\%} hit the coverage ceiling and just \num{\revisionTwo{12.4}}{\%} contain test deletions, yet \num{\revisionTwo{74.8}}{\%} add new tests that nonetheless fail to cover the agent's own changed lines\Space{, suggesting tests that exercise neighboring rather than newly introduced code}.\Space{
\Fix{Percentages in this paragraph weirdly shift to no decimal sig fig. Should continue to use 1 decimal sig fig.}}

\noindent\textbf{Takeaway.} \revision{Agent-written tests significantly improve coverage on average, but only a minority of PRs drive this gain: just \JavaOverallPctImp{}\% of Java and \PythonOverallPctImp{}\% of Python Code\,+\,Tests PRs add coverage beyond the existing suite. Failures look different by language: Java agents often delete tests or hit a coverage ceiling, while Python agents add tests that exercise code other than the lines they introduced.}

\noindent\textbf{Categories of code that agents leave untested.} Table~\ref{tab:miss-category-side-by-side} reports miss rates by syntactic category for the added executable lines\Space{ (Section~\ref{sec:artifact-gen})}. Two patterns dominate.

First, the cross-language gap in overall diff coverage extends to almost every category: Python miss rates run roughly 2--6$\times$ higher than Java's across \texttt{\textbf{\small Method Call}}, \texttt{\textbf{\small Assignment}}, \texttt{\textbf{\small If}}, \texttt{\textbf{\small For}}, and \texttt{\textbf{\small Return}}. Even common constructs like \texttt{\textbf{\small Assignment}} (\JavaMissCatAssignmentMiss\% Java vs. \PythonMissCatAssignmentMiss\% Python) and \texttt{\textbf{\small Method Call}} (\JavaMissCatMethodCallMiss\% vs. \PythonMissCatMethodCallMiss\%) are largely untested in Python.

Second, error-handling lines are under-tested in \emph{both} languages. Newly added \texttt{\textbf{\small Throw}} statements miss \JavaMissCatThrowMiss\% in Java and \PythonMissCatThrowMiss\% of the time in Python; lines inside \texttt{Try-Catch} blocks miss \JavaMissCatTryCatchMiss\% and \PythonMissCatTryCatchMiss\%, respectively. Notably, Java's \texttt{\textbf{\small Try-Catch}} miss rate is the only category where Java performs \emph{worse} than Python in absolute terms, despite its substantially higher overall coverage.

\noindent\textbf{Takeaway.} Agent test misses are not uniform across syntax. Error handling is the weakest spot in both languages, with most \texttt{\textbf{\small Throw}} statements and \texttt{\textbf{\small Try-Catch}} bodies left unexercised. In Python, the gap extends to routine constructs: \texttt{\textbf{\small Assignment}}, \texttt{\textbf{\small Method Call}}, and \texttt{\textbf{\small Return}} are all missed more than half the time—suggesting agents exercise happy paths, while leaving large portions of their own diff unexercised.

\begin{table}[t]
\centering
\caption{Diff-coverage miss counts by statement category, compared between Python and Java. \textit{Total} is the number of executable lines of that category added by agents; \textit{Miss\%} is the fraction of those lines that no tests can cover.}
\label{tab:miss-category-side-by-side}
\vspace{-1em}
\begin{tabular}{l rr rr}
\toprule
& \multicolumn{2}{c}{\textbf{Python}} & \multicolumn{2}{c}{\textbf{Java}} \\
\cmidrule(lr){2-3} \cmidrule(lr){4-5}
\textbf{Category} & \textbf{Total} & \textbf{Miss\%} & \textbf{Total} & \textbf{Miss\%} \\
\midrule
Method Call & \PythonMissCatMethodCallTotal & \PythonMissCatMethodCallMiss\% & \JavaMissCatMethodCallTotal & \JavaMissCatMethodCallMiss\% \\
Assignment & \PythonMissCatAssignmentTotal & \PythonMissCatAssignmentMiss\% & \JavaMissCatAssignmentTotal & \JavaMissCatAssignmentMiss\% \\
If & \PythonMissCatIfTotal & \PythonMissCatIfMiss\% & \JavaMissCatIfTotal & \JavaMissCatIfMiss\% \\
For & \PythonMissCatForTotal & \PythonMissCatForMiss\% & \JavaMissCatForTotal & \JavaMissCatForMiss\% \\
While & \PythonMissCatWhileTotal & \PythonMissCatWhileMiss\% & \JavaMissCatWhileTotal & \JavaMissCatWhileMiss\% \\
Return & \PythonMissCatReturnTotal & \PythonMissCatReturnMiss\% & \JavaMissCatReturnTotal & \JavaMissCatReturnMiss\% \\
Try-Catch & \PythonMissCatTryCatchTotal & \PythonMissCatTryCatchMiss\% & \JavaMissCatTryCatchTotal & \JavaMissCatTryCatchMiss\% \\
Throw & \PythonMissCatThrowTotal & \PythonMissCatThrowMiss\% & \JavaMissCatThrowTotal & \JavaMissCatThrowMiss\% \\
Break & \PythonMissCatBreakTotal & \PythonMissCatBreakMiss\% & \JavaMissCatBreakTotal & \JavaMissCatBreakMiss\% \\
Continue & \PythonMissCatContinueTotal & \PythonMissCatContinueMiss\% & \JavaMissCatContinueTotal & \JavaMissCatContinueMiss\% \\
Switch & \PythonMissCatSwitchTotal & \PythonMissCatSwitchMiss\% & \JavaMissCatSwitchTotal & \JavaMissCatSwitchMiss\% \\
Definition & \PythonMissCatDefinitionTotal & \PythonMissCatDefinitionMiss\% & \JavaMissCatDefinitionTotal & \JavaMissCatDefinitionMiss\% \\
Import & \PythonMissCatImportTotal & \PythonMissCatImportMiss\% & — & — \\
Other & \PythonMissCatOtherTotal & \PythonMissCatOtherMiss\% & \JavaMissCatOtherTotal & \JavaMissCatOtherMiss\% \\
\bottomrule
\end{tabular}
\vspace{-1.5em}
\end{table}

\section{Discussion and Future Direction}
\label{sec:discussion}

\textbf{For practitioners.} \revision{Do not assume agents write tests\Space{ (Finding i)}: \num{50.4}\% of \CodeUnderTest{}-modifying PRs include no test changes at all, and the safety net provided by existing tests is incomplete in both languages\Space{ (Finding ii)}. Existing tests cover only \JavaExistingCov{}\% of agents' changed executable lines in Java and \PythonExistingCov\% in Python. with existing tests covering no changed lines in \num{64.8}\% of Python PRs. Teams using agentic PRs should not assume a passing run of tests means that the change has been tested. Review efforts should prioritize error-handling paths\Space{ (Finding iv)} \revisionTwo{as \num{\JavaMissCatTryCatchMiss}\% Java and \num{\PythonMissCatTryCatchMiss}\% Python\Space{\Fix{This is outdated and wrong? Macro all numbers PLEASE.}} of the them} are not covered regardless of whether the agent added tests.}

\textbf{For agent developers.} \revision{Agents that do include tests still fail to test their own changes most of the time\Space{ (Finding iii)}. Agent-written tests improve coverage on average, but add coverage of the agent's own changes in only 35.9\% of Java and 22.5\% of Python Code + Tests PRs. The two languages fail differently: Java agents sometimes \emph{delete} more tests than they add (a \num{2.6}$\times$ deletion-to-addition ratio in non-improving Java PRs), while Python agents add tests that exercise some other code than the code they introduced. Both patterns point to the same intervention: agents need a coverage-aware feedback loop that checks whether their own added lines are exercised by their own added tests before submitting a PR. As existing tests often leave agent changes not covered\Space{ (Finding ii)}, agents cannot assume existing tests will test their newly added code.  Error-handling constructs (e.g., \texttt{throw}/\texttt{raise}, \texttt{try}/\texttt{catch}) should be the priority target\Space{ (Finding iv)} for agent code test generation as \revisionTwo{\num{\JavaMissCatTryCatchMiss}\% Java and \num{\PythonMissCatTryCatchMiss}\% Python}\Space{\Fix{same issue as last paragraph}} of them are not covered.}

\textbf{For researchers.} Our findings spur key new research:

\textit{(1) Coverage-aware agent benchmarks\Space{ (Findings i, ii)}}. SWE-bench~\cite{jimenez2024swebench} and similar benchmarks reward agents for passing existing tests, but do not analyze whether agents test the code they introduce. Future benchmarks should focus on diff coverage\Space{, reported by language}.

\textit{(2) Diagnose why agent tests miss the lines they introduce\Space{ (Finding iii)}}. For \emph{Code\,+\,Tests} PRs that show no diff-coverage gain, our current analysis tells us that the gain is absent but not \emph{why}. Diagnosing this behavior is an open problem whose answer can inform agent design and benchmark construction.

\textit{(3) Targeted test-generation for weak categories\Space{ (Finding iv)}}. Error-handling constructs miss \revisionTwo{\num{\JavaMissCatTryCatchMiss}\% in Java and \num{\PythonMissCatTryCatchMiss}\% in Python}\Space{ \Fix{same as before}}. How can targeted prompting shift agents' attention toward such constructs remains an important future work direction.

\section{Related Work, Limitations}

\textbf{Related work.} One existing body of work evaluates LLMs as test generators under controlled conditions. Pizzorno and Berger~\cite{lemieux2023coverup} and Ryan et al.~\cite{ryan2024code} used coverage-guided prompting to drive LLM-generated tests toward higher branch and line coverage; Meta's TestGen-LLM~\cite{meta2024testgen} reported code coverage gains\Space{ \Fix{What is production gains? code coverage gains?}} from augmenting existing tests with LLM-written tests; El Haji et al.~\cite{schafer2024empirical} evaluated Copilot for Python test generation; and Mastropaolo et al.~\cite{mastropaolo2023robustness} studied the robustness of Copilot's code generation more broadly. 
These studies established that LLMs \emph{can} produce useful tests when explicitly asked, but they evaluated models in a test-generation role rather than observing what agents do when given autonomy over a PR. 
Watanabe et al.~\cite{watanabe2025use} analyzed \num{567} agentic PRs and reported that \num{45.1}\% of merged PRs required human revisions, but do not measure test coverage. Coverage-based benchmarks such as SWE-bench~\cite{jimenez2024swebench} evaluate whether agents \emph{pass} curated tests, treating the tests as fixed and externally provided, but none measure whether existing tests or tests an agent adds actually exercise the code that the agent introduced.

\textbf{Limitations}. Our coverage analysis is limited to merged PRs
from repositories that could be built and instrumented, which may introduce sampling bias. To assess this threat, we compared lines of code (LOC), number of commits, age, stars, and forks between the analyzed repositories and the entire set of projects (median of analyzed vs.\ entire set). The \num{10} analyzed Java repositories are substantially smaller (\num{29}K vs.\ \num{102}K LOC), less starred (\num{437} vs.\ \num{745}), and less forked (\num{116} vs.\ \num{254}) than the entire set of Java projects. 
That being said, the analyzed projects are comparable in number of commits (\num{2271} vs.\ \num{3983} commits) and median age (\num{4178} vs.\ \num{3785} days). 
Our Java coverage results may thus primarily reflect the testing practices of smaller projects. 
In contrast, the \num{34} analyzed Python repositories show no substantial differences from the entire set of Python projects on any metric (\num{31}K vs.\ \num{25}K LOC, \num{1202} vs.\ \num{872} commits, \num{1065} vs.\ \num{1268} days, \num{676} vs.\ \num{734} stars, \num{120} vs.\ \num{141} forks).
Full statistics and distribution plots are in our replication package.\Space{\Fix{Given how we have some space now, should we include some statistics about the distribution of what we ran compared to the whole dataset? E.g., avg LOC of what we analyzed and the full set of projects, avg. stars, ...}}

\section{Conclusions}
\revision{In this paper, we performed a cross-language study of how well tested agentic PRs are in the wild, which our results show are often poorly tested. 
Over half of \CodeUnderTest{}-changing PRs include no test changes; existing tests cover only \JavaExistingCov\% of agents' changed lines in Java and \PythonExistingCov\% in Python; agent-written tests improve coverage, but only in a minority of PRs; and error-handling constructs are often missed. Our findings spur new research\Space{ directions} to better help test agent generated code.}

\Space{\Fix{Are [3] and [4] both books? If so, they should have the same information. Does [4] have authors? [1] capitalizes every word but [2] does not. [7] has page numbers repeated twice. [11] also capitalizes all words unlike most other references. Ditto for [12], [13], [15], [26]. [17] should have doi. see comment. [17] and [12] are both articles but they contain different info. [18] is a conference citation that has vol and missing page numbers unlike other conference citations. pytest and JaCoCo can drop or change the 2025/2024 to 2026. [22] should drop the ``Initial Release'' part. AI in [16] should be capitalized. [13] is misleadingly shown as if it is a research track publication when it should be stated as Industry Track. See comment for example. Is it SWT-Bench or SWT-bench? On the website it seems to be SWT-bench.}}
\Comment{https://dl.acm.org/doi/10.1145/157710.157715}
\Comment{[20] in https://people.cs.gmu.edu/~winglam/publications/2025/RahmanETAL25RankF.pdf}

\balance
\bibliographystyle{IEEEtran}
\bibliography{references}

@string{ESEC/FSE = "European Software Engineering Conference and Symposium on the Foundations of Software Engineering"}

@string{FSEIndustry = "Symposium on the Foundations of Software Engineering, industry track"}

@string{ICLR = "International Conference on Learning Representations"}

@string{ICSE = "International Conference on Software Engineering"}

@string{ICSM = "International Conference on Software Maintenance"}

@string{NeurIPS = "Advances in Neural Information Processing Systems"}

@string{PACMSE = "Proceedings of the ACM on Software Engineering"}

@string{TOSEM = "ACM Transactions on Software Engineering Methodology"}

@string{Software = "IEEE Software"}

@string{AST = "International Conference on Automation of Software Test"}

@string{IST = "Information and Software Technology"}

@string{ESE = "Empirical Software Engineering"}

@string{SPE = "Software: Practice and Experience"}

@misc{pytest-cov,
  title        = {{pytest-cov}: Coverage Plugin for {pytest}},
  author       = {{pytest-dev contributors}},
  year=2026,
  howpublished = {\url{https://pypi.org/project/pytest-cov/}}
}

@article{chen2021evaluating,
  author    = {Chen, Mark and Tworek, Jerry and Jun, Heewoo and Yuan, Qiming and Pinto, Henrique Ponde de Oliveira and Kaplan, Jared and Edwards, Harri and Burda, Yuri and Joseph, Nicholas and Brockman, Greg and Ray, Alex and Puri, Raul and Krueger, Gretchen and Petrov, Michael and Khlaaf, Heidy and Sastry, Girish and Mishkin, Pamela and Chan, Brooke and Gray, Scott and Ryder, Nick and Pavlov, Mikhail and Power, Alethea and Kaiser, Lukasz and Bavarian, Mohammad and Winter, Clemens and Tillet, Philippe and Such, Felipe Petroski and Cummings, Dave and Plappert, Matthias and Chanez, Fotios and Barnes, Elizabeth and Herbert-Voss, Ariel and Guss, William Hebgen and Nichol, Alex and Paino, Alex and Tezak, Nikolas and Tang, Jie and Babuschkin, Igor and Balaji, Suchir and Jain, Shantanu and Saunders, William and Hesse, Christopher and Carr, Andrew N. and Leike, Jan and Achiam, Josh and Misra, Vedant and Morikawa, Evan and Radford, Alec and Knight, Matthew and Brundage, Miles and Murati, Mira and Mayer, Katie and Welinder, Peter and McGrew, Bob and Amodei, Dario and McCandlish, Sam and Sutskever, Ilya and Zaremba, Wojciech},
  title     = {Evaluating Large Language Models Trained on Code},
  journal   = {arXiv preprint arXiv:2107.03374},
  year      = {2021},
  url={https://doi.org/10.48550/arXiv.2107.03374},
doi={10.48550/arXiv.2107.03374}
}

@inproceedings{mundler2024swtbench,
  title={{SWT}-Bench: Testing and validating real-world bug-fixes with code agents},
  author={M{\"u}ndler, Niels and M{\"u}ller, Mark N and He, Jingxuan and Vechev, Martin},
  booktitle=NeurIPS,
  pages = {81857-81887},
  year={2024},
  url={https://doi.org/10.52202/079017-2601}
}

@book{ammann2016introduction,
  author    = {Ammann, Paul and Offutt, Jeff},
  title     = {Introduction to Software Testing},
  publisher = {Cambridge University Press},
  year      = {2016}
}

@book{pezze2007software,
  title={Software testing and analysis: process, principles, and techniques},
  author={Pezz{\`e}, Mauro and Young, Michal},
  year={2008},
  publisher={John Wiley \& Sons}
}

@article{graves2001empirical,
  author    = {Graves, Todd L. and Harrold, Mary Jean and Kim, Jung-Min and Porter, Adam and Rothermel, Gregg},
  title     = {An Empirical Study of Regression Test Selection Techniques},
  journal   = TOSEM,
  volume    = {10},
  pages    = {184--208},
  year      = {2001},
  url       = {https://doi.org/10.1145/367008.367020}
}

@ARTICLE{swebok2014,
  author={Bourque, P. and Dupuis, R. and Abran, A. and Moore, J.W. and Tripp, L.},
  journal=Software, 
  title={The guide to the Software Engineering Body of Knowledge}, 
  year={1999},
  volume={16},
  pages={35--44},
  url={https://doi.org/10.1109/52.805471}
}

@article{pecorelli2021relation,
  title={The relation of test-related factors to software quality: A case study on {A}pache systems},
  author={Pecorelli, Fabiano and Palomba, Fabio and De Lucia, Andrea and Bacchelli, Alberto},
  journal=ESE,
  volume={26},
  pages={1--42},
  year={2021},
  url={https://doi.org/10.1007/s10664-020-09891-y}
}

@article{bhatia2024empirical,
author = {Bhatia, Aaditya and Khomh, Foutse and Adams, Bram and Hassan, Ahmed},
title = {An Empirical Study of Self-Admitted Technical Debt in Machine Learning Software},
year = {2026},
volume = {35},
pages = {1--44},
url = {https://doi.org/10.1145/3785001},
journal = TOSEM
}

@article{cunningham1992wycash,
  author={Cunningham, Ward},
  title={The {WyCash} portfolio management system},
  journal = {ACM SIGPLAN OOPS Messenger},
  volume = {4},
  pages={29--30},
  year={1992},
  url = "https://dl.acm.org/doi/10.1145/157709.157715"
}

@article{ryan2024code,
  author    = {Gabriel Ryan and Siddhartha Jain and Mingyue Shang and Shiqi Wang and Xiaofei Ma and Murali Krishna Ramanathan and Baishakhi Ray},
  title     = {Code-Aware Prompting: A Study of Coverage-Guided Test Generation in Regression Setting using {LLM}},
  journal   = {Proceedings of the ACM on Software Engineering},
  volume    = {1},
  pages = {951--971},
  year      = {2024},
  url={https://doi.org/10.1145/3643769}
}

@inproceedings{sculley2015hidden,
  title={Hidden technical debt in machine learning systems},
  author={Sculley, D. and Holt, Gary and Golovin, Daniel and Davydov, Eugene and Phillips, Todd and Ebner, Dietmar and Chaudhary, Vinay and Young, Michael and Crespo, Jean-Fran{\c{c}}ois and Dennison, Dan},
  booktitle=NeurIPS,
  year={2015}
}

@misc{treesitter2018,
  author = {Brunsfeld, Max},
  title = {Tree-sitter: An Incremental Parsing System for Programming Tools},
  howpublished = {\url{https://tree-sitter.github.io/tree-sitter/}},
  note = {Accessed: 2026}
}

@inproceedings{anonymous2026aiprtest,
  title     = {Replication package},
  author    = {SageSELab},
  booktitle = {IEEE International Conference on Software Maintenance and Evolution (ICSME)},
  year      = {2026},
  note      = {Replication package: \url{https://github.com/SageSELab/Agentic-Pull-Request-Test-Coverage/}}
}

@software{dipongkor2026artifact,
  author    = {Dipongkor, Atish Kumar},
  title     = {{S}age{SEL}ab/{A}gentic-{P}ull-{R}equest-{T}est-{C}overage},
  year      = {2026},
  month     = jul,
  publisher = {Zenodo},
  version   = {1.0.0},
  doi       = {10.5281/zenodo.21419686},
  url       = {https://doi.org/10.5281/zenodo.21419686}
}

@misc{openai2021codex,
  author       = {{OpenAI}},
  title        = {{OpenAI Codex}},
  year         = {2021},
  howpublished = {\url{https://openai.com/blog/openai-codex}},
  note         = {Accessed: 2025}
}

@misc{github2022copilot,
  author       = {{GitHub}},
  title        = {{GitHub Copilot}: Your {AI} Pair Programmer},
  year         = {2022},
  howpublished = {\url{https://github.com/features/copilot}},
  note         = {Accessed: 2025}
}

@misc{cursor2024,
  author       = {{Anysphere}},
  title        = {{Cursor}: The {AI} Code Editor},
  year         = {2024},
  howpublished = {\url{https://cursor.com}},
  note         = {Accessed: 2025}
}

@misc{anthropic2025claudecode,
  author       = {{Anthropic}},
  title        = {{Claude Code}},
  year         = {2025},
  howpublished = {\url{https://www.anthropic.com/claude-code}},
  note         = {Accessed: 2025}
}

@article{li2025aidev,
title={The Rise of {AI} Teammates in Software Engineering {(SE)} 3.0: How Autonomous Coding Agents Are Reshaping Software Engineering}, 
author={Li, Hao and Zhang, Haoxiang and Hassan, Ahmed E.},
journal={arXiv preprint arXiv:2507.15003},
year={2025},
url={https://doi.org/10.48550/arXiv.2507.15003},
doi={10.48550/arXiv.2507.15003}
}

@misc{jacoco,
  author       = {{EclEmma Team}},
  title        = {{JaCoCo}: {Java} Code Coverage Library},
  year         = {2026},
  howpublished = {\url{https://www.jacoco.org/jacoco/}}
}

@misc{cognition2024devin,
  author       = {{Cognition Labs}},
  title        = {Introducing Devin, the First {AI} Software Engineer},
  year         = {2024},
  howpublished = {\url{https://cognition.ai/blog/introducing-devin}},
  note         = {Accessed: 2025}
}

@inproceedings{fraser2011evosuite,
  author    = {Gordon Fraser and Andrea Arcuri},
  title     = {{EvoSuite}: Automatic Test Suite Generation for Object-Oriented Software},
  booktitle = ESEC/FSE,
  pages     = {416--419},
  year      = {2011},
  url = {https://doi.org/10.1145/2025113.2025179},
}

@inproceedings{mastropaolo2023robustness,
  author    = {Antonio Mastropaolo and Luca Pascarella and Emanuela Guglielmi and Matteo Ciniselli and Simone Scalabrino and Rocco Oliveto and Gabriele Bavota},
  title     = {On the Robustness of Code Generation Techniques: An Empirical Study on {GitHub Copilot}},
  booktitle = ICSE,
  pages     = {2149--2160},
  year      = {2023},
  url={https://doi.org/10.1109/ICSE48619.2023.00181}
}

@inproceedings{schafer2024empirical,
  title={Using {G}it{H}ub {C}opilot for test generation in {P}ython: An empirical study},
  author={El Haji, Khalid and Brandt, Carolin and Zaidman, Andy},
  booktitle = AST,
  pages={45--55},
  year={2024},
  url={https://doi.org/10.1145/3644032.3644443},
  doi={10.1145/3644032.3644443}
}

@article{lemieux2023coverup,
author = {Altmayer Pizzorno, Juan and Berger, Emery D.},
title = {Cover{U}p: Effective High Coverage Test Generation for {P}ython},
journal = PACMSE,
year = {2025},
volume = {2},
pages={2897--2919},
url = {https://doi.org/10.1145/3729398}
}

@inproceedings{meta2024testgen,
author = {Alshahwan, Nadia and Chheda, Jubin and Finogenova, Anastasia and Gokkaya, Beliz and Harman, Mark and Harper, Inna and Marginean, Alexandru and Sengupta, Shubho and Wang, Eddy},
title = {Automated Unit Test Improvement using Large Language Models at {M}eta},
year = {2024},
url = {https://doi.org/10.1145/3663529.3663839},
booktitle = FSEIndustry,
pages = {185--196}
}

@inproceedings{
    jimenez2024swebench,
    title={{SWE}-bench: Can Language Models Resolve Real-world Github Issues?},
    author={Carlos E Jimenez and John Yang and Alexander Wettig and Shunyu Yao and Kexin Pei and Ofir Press and Karthik R Narasimhan},
    booktitle=ICLR,
    year={2024},
    url={https://openreview.net/forum?id=VTF8yNQM66}
}

@article{watanabe2025use,
author = {Watanabe, Miku and Li, Hao and Kashiwa, Yutaro and Reid, Brittany and Iida, Hajimu and Hassan, Ahmed E.},
title = {On the Use of Agentic Coding: An Empirical Study of Pull Requests on {G}it{H}ub},
year = {2026},
url = {https://doi.org/10.1145/3798166},
journal = TOSEM
}

@article{liang2024ai_assistants,
  title={Using {AI}-based coding assistants in practice: State of affairs, perceptions, and ways forward},
  author={Agnia Sergeyuk and Yaroslav Golubev and Timofey Bryksin and Iftekhar Ahmed},
  journal = IST,
  volume={178},
  year={2025},
  pages = {107610},
  url={https://doi.org/10.1016/j.infsof.2024.107610}
}

@article{zhu2015analysis,
  title={An analysis of programming language statement frequency in {C}, {C}++, and {J}ava source code},
  author={Zhu, Xiaoyan and Whitehead Jr, E James and Sadowski, Caitlin and Song, Qinbao},
  journal=SPE,
  volume={45},
  number={11},
  year={2015},
  url={https://doi.org/10.1002/spe.2298}
}

@incollection{wilcoxon1992individual,
  title={Individual comparisons by ranking methods},
  author={Wilcoxon, Frank},
  booktitle={Breakthroughs in statistics: Methodology and distribution},
  pages={196--202},
  year={1992},
  url={https://doi.org/10.1007/978-1-4612-4380-9_16}
}

@INPROCEEDINGS{collard2013srcml,
  author={Collard, Michael L. and Decker, Michael John and Maletic, Jonathan I.},
  booktitle=ICSM, 
  title={src{ML}: An Infrastructure for the Exploration, Analysis, and Manipulation of Source Code: A Tool Demonstration}, 
  year={2013},
  pages={516-519},
  url={https://doi.org/10.1109/ICSM.2013.85}
}

\end{document}